\documentclass[prd,amsmath,amssymb,twocolumn]{revtex4}
\usepackage{hyperref,amssymb,amsbsy}
\pdfoutput=1

\newcommand{\curl}{\,\mbox{curl}\,}

\begin{document}
\title{First-order adiabatic perturbations of a perfect fluid about a general FLRW background using the $\mathbf{1+3}$ covariant and gauge-invariant formalism}
\author{Sylvain D. Brechet}
\email{sdb41@mrao.cam.ac.uk}
\author{Michael P. Hobson}
\email{mph@mrao.cam.ac.uk}
\author{Anthony N. Lasenby}
\email{a.n.lasenby@mrao.cam.ac.uk}
\affiliation{Astrophysics Group, Cavendish Laboratory, J.~J.~Thomson Avenue, Cambridge, CB3 0HE, UK}
\begin{abstract}
An analysis of adiabatic perturbations of a perfect fluid is performed to first-order about a general FLRW background using the $1+3$ covariant and gauge-invariant formalism. The analog of the Mukhanov-Sasaki variable and the canonical variables needed to quantise respectively the scalar and tensor perturbations in a general FLRW background space-time are identified. The dynamics of the vector perturbations is also discussed. 
\end{abstract}
\maketitle

\section{Introduction}

Cosmological perturbation theory about a given background metric was initiated by Lifshitz~\cite{Lifshitz:1946} and extended notably by Bardeen~\cite{Bardeen:1980}, Mukhanov, Feldmann \& Brandenberger~\cite{Mukhanov:1992}, and Kodama \& Sasaki~\cite{Sasaki:1984}. This is the standard approach to performing a perturbation analysis of general relativity in order to describe the `realistic' dynamics of cosmological models. It has the disadvantage that truly physical results can be obtained only after completely specifying the correspondence between the `real' perturbed space-time and a `fictitious' unperturbed background space-time, which is chosen to be highly symmetric. Such a correspondence is not uniquely defined and changes under a gauge transformation. The degrees of freedom in the definition of the correspondence, or the gauge freedom, leave unphysical gauge modes in the dynamical equations describing the evolution of the perturbations~\cite{Bruni:1992}. This gauge problem~\cite{Lifshitz:1963, Sachs:1967} is inherent to such a `background-based' perturbation approach. Indeed, the metric, and consequently the Einstein equations, have $10$ degrees of freedom whereas the dynamics is determined by $6$ parameters only. Therefore, the $4$ remaining degrees of freedom are directly related to the gauge.

To solve this problem, Bardeen~\cite{Bardeen:1980} determined a set of gauge-invariant quantities to describe the perturbations and derived their dynamical equations. These quantities are mathematically well defined but do not have a transparent geometrical meaning since they are defined with respect to a particular coordinate system~\cite{Stewart:1990}, and their physical meaning is obscure~\cite{Ellis:1989b}.

An alternative way to circumvent the gauge problem is to follow the $1+3$ covariant approach, which was developed by Hawking~\cite{Hawking:1966} and extended by Olsen~\cite{Olson:1976} and Ellis \& Bruni~\cite{Ellis:1989}. The aim of this approach is to study the dynamics of real cosmological fluid models in a physically transparent manner. This formalism relies on covariantly defined variables, which are gauge-invariant by construction~\cite{Bruni:1992}, thus simplifying the methodology and clarifying the physical interpretation of the models. It also allows the metric to be arbitrary. This approach admits a covariant and gauge-invariant linearisation that allows a perturbation analysis to be performed in a direct manner~\cite{Challinor:2000}.

The basic philosophy of the perturbation theory based on the $1+3$ covariant formalism is different from the `background-based' perturbation theory~\cite{Ellis:1990}. Instead of starting with a background space-time and then perturbing it, the approach begins with an inhomogeneous and anisotropic (real) space-time, which reduces to the background space-time on large scales. In our case, we will take the background to be a homogeneous and isotropic FLRW space-time, although this formalism allows in principle for more complicated backgrounds. Therefore, the `real' space-time has been appropriately called an almost FLRW space-time~\cite{Ellis:1989b}. In this perturbation theory, the approximation takes place by neglecting higher-order terms in the exact equations when the values of the kinematic and dynamic variables are close to those they would take in the background FLRW space-time. The analysis is performed in the real space-time and the dynamical equations are subsequently linearised. The background solution is simply the zero-order approximation of the exact solution of the dynamical equations.

Although in the $1+3$ covariant formalism, the scalar, vector and tensor perturbations are handled in a unified way~\cite{Challinor:2000}, they decouple to first-order and can be studied independently~\cite{Mukhanov:1992}. For a perfect fluid in a spatially-curved case, the Bardeen variable for the scalar perturbations in the $1+3$ covariant approach was first identified by Woszczyna \& Kulak~\cite{Woszczyna:1989} and the curvature perturbation by Bruni, Dunsby \& Ellis~\cite{Bruni:1992}. The first attempt to establish an explicit relation between the perturbation variables in the $1+3$ covariant formalism and the corresponding variables in the `background-based' approach was made by Goode~\cite{Goode:1989}.

In the $1+3$ formalism, the time derivative of a physical quantity is defined usually as the projection of the covariant derivative of the quantity on the worldline as outlined in Sec.~\ref{Kinematics in the 1+3 covariant formalism}, but this is not the only way to define covariantly a time derivative as Thiffeault showed in~\cite{Thiffeault:2001}. As Langlois \& Vernizzi suggested~\cite{Langlois:2005}, the Lie derivative along the worldline of a fluid element is another possible definition. However, the Lie derivative of a scalar field along the worldline is identical to time derivative of the scalar field along the worldline. Since, in this publication, we aim to identify scalar quantities, which are respectively scalar perturbations and the scalar amplitude of vector and tensor perturbations, we will define the time derivative of physical quantities to be the projection of the covariant derivative of the quantity on the worldline.

It is worth mentioning that, recently, Pitrou \& Uzan~\cite{Pitrou:2007}, working in a spatially-flat case, used Lie derivatives to recast the dynamical equations for the scalar and tensor perturbations in the $1+3$ covariant formalism in order to identify perturbation variables which are similar to the Sasaki-Mukhanov and canonical variables. In particular, they sought to identify the scalar and tensor variables that map to the Mukhanov-Sasaki and to the tensor canonical variables when considering a spatially-flat almost FLRW universe. In the current work, we seek the extension to general FLRW universes. However, there are some further differences with the work of Pitrou \& Uzan. Firstly, in the scalar case, the perturbation variable $v_a$ that Pitrou \& Uzan obtained is not a scalar but a covector. We will seek a scalar here. Secondly, for tensor perturbations in the spatially-flat case, Pitrou \& Uzan obtained a wave equation in terms of Lie derivatives of tensor fields. In the current work, we will extend their analysis by seeking to identify canonical variables, which correspond to the scalar amplitudes of the tensor perturbations. These scalar amplitudes are the canonical variables defined by Grishchuk~\cite{Grishchuk:1974}. We will be using the name `canonical' to refer to these variables henceforth in this publication. Finally, Pitrou \& Uzan did not consider a perfect fluid analysis as we carry out here, but instead restricted their analysis to a single scalar field.

Thus the aim of this publication is to identify, in the $1+3$ covariant approach, for adiabatic perturbations of a perfect fluid in spatially-curved FLRW models, the analog of the Mukhanov-Sasaki variable~\cite{Mukhanov:1992, Sasaki:1984} and the canonical  variables~\cite{Grishchuk:1974} needed to quantise respectively the scalar and tensor perturbations. The quantisation of fields in a spatially-curved space-time lies outside the scope of this paper. However, this topic has been addressed in detail by Birrell \& Davis~\cite{Birrell:1982} and Fulling~\cite{Fulling:1989}. Our approach broadens the scalar perturbations analysis performed by Woszczyna \& Kulak~\cite{Woszczyna:1989}, Bruni, Dunsby \& Ellis~\cite{Bruni:1992} and Lyth \& Woszczyna~\cite{Lyth:1995} in a spatially-curved case.         

The structure of this publication is as follows. In Sec.~\ref{Kinematics in the 1+3 covariant formalism}, we give a concise description of the kinematics in the $1+3$ covariant approach. In Sec.~\ref{First-order adiabatic dynamics of a perfect fluid about an FRW background}, we establish the first-order dynamical equations for adiabatic perturbations of a perfect fluid in the 1+3 covariant approach. In Sec.~\ref{Scalar perturbations}, we identify the analog of the Mukhanov-Sasaki variable for the scalar perturbations. In Sec.~\ref{Vector perturbations}, we determine the dynamics of the scalar amplitude of the vector perturbations. In Sec.~\ref{Tensor perturbations}, we finally identify the canonical variables for the tensor perturbations.

For convenience, we follow Hawking~\cite{Hawking:1966} and Ellis~\cite{Ellis:1989} by adopting the $(-,+,+,+)$-signature for the metric. This choice of signature is particularly appropriate for the $1+3$ covariant formalism. Indeed, the dynamical equations are projected on the local spatial hypersurfaces, which are positively defined for our signature convention. The correspondence for dynamical quantities expressed in terms of the opposite signature can be found in~\cite{Brechet:2007}.

\section{Kinematics in the 1+3 covariant formalism}
\label{Kinematics in the 1+3 covariant formalism}

We will briefly outline the basics of the $1+3$ covariant formalism describing the fluid kinematics. To introduce this formalism, we follow Ellis \& van Elst's approach~\cite{Ellis:1999}. The approach is based on the $1+3$ decomposition of geometric quantities with respect to a fundamental $4$-velocity $u^{a}$ that uniquely determines the frame and the worldline of every infinitesimal volume element of fluid,
\begin{equation}
u^{a}=\frac{dx^{a}}{d\tau}\ ,\ \ \ \ \ \ \ u_{a}u^{a}=-1\ ,
\end{equation}
where $x^{a}$ are arbitrary cosmic coordinates, and $\tau$ is the proper time measured along the worldlines. In the context of a general cosmological model, we require that the $4$-velocity be chosen in a physical manner such that in the FLRW limit the dipole of the cosmic microwave background radiation vanishes. This condition is necessary to ensure the gauge-invariance of the approach.

The $4$-velocity $u^{a}$ defines locally and in a unique fashion two projection tensors,
\begin{equation}
\begin{split}
&U_{ab} = -u_{a}u_{b}\ ,\\
&h_{ab} = g_{ab}+u_{a}u_{b}\ ,
\end{split}
\end{equation}
which satisfy the following properties,
\begin{equation}
\begin{split}
&{U^{a}}_{c}{U^{c}}_{b} = {U^{a}}_{b} \ , \ \ {U^{a}}_{a} = 1 \ , \ \ U_{ab}u^{b} = u_{a}\ ,\\
&{h^{a}}_{c}{h^{c}}_{b} = {h^{a}}_{b} \ , \ \ \ \ {h^{a}}_{a} = 3 \ , \ \, \ h_{ab}u^{b} = 0\ .
\end{split}
\end{equation}

The first projects parallel to the 4-velocity vector $u^{a}$, and the second determines the metric properties of the instantaneous rest frame of the fluid. It is useful to introduce the totally antisymmetric Levi-Civita tensor $\varepsilon_{abc}$, which is defined in the rest frame of the fluid and thus satisfies,
\begin{equation}
\varepsilon_{abc}u^{c} = 0\ .
\end{equation}

Moreover, we define two projected covariant derivatives which are the time projected covariant derivative along the worldline (denoted $\mathbf{\dot{}}\ $) and the spatially projected covariant derivative (denoted $D_{a}$). For any quantity ${Q^{a\dots}}_{b\dots}$, these are respectively defined as
\begin{equation}
\begin{split}
&{{\dot Q}^{a\dots}}_{\ \ \ \, b\dots}\equiv u^{c}\nabla_{c}{Q^{a\dots}}_{b\dots}\ ,\label{time covariant}\\
&D_{c}{Q^{a\dots}}_{b\dots}\equiv {h^{f}}_{c}{h^{a}}_{d}\dots {h^{e}}_{b}\dots\nabla_{f}{Q^{d\dots}}_{e\dots}\ .
\end{split}
\end{equation}
Furthermore, the kinematics and the dynamics are determined by projected tensors that are orthogonal to $u^{a}$ on every index. The angle brackets are used to denote respectively orthogonal projections of vectors $V^{a}$ and the orthogonally projected symmetric trace-free part $(\mathrm{PSTF})$ of rank-$2$ tensors $T^{ab}$ according to,
\begin{equation}
\begin{split}
&V^{\langle a\rangle} ={h^{a}}_{b}V^{b}\ ,\\
&T^{\langle ab\rangle} =\left({h^{(a\vphantom)}}_{c}{h^{\vphantom(b)}}_{d}-{\textstyle\frac{1}{3}}h^{ab}h_{cd}\right)T^{cd}\ .
\end{split}
\end{equation}
For convenience, the angle brackets are also used to denote the orthogonal projections of covariant time derivatives of vectors and tensors along the worldline $u^{a}$ as follows,
\begin{equation}
\begin{split}
&\dot{V}^{\langle a\rangle} ={h^{a}}_{b}\dot{V}^{b}\ ,\\
&\dot{T}^{\langle ab\rangle} =\left({h^{(a\vphantom)}}_{c}{h^{\vphantom(b)}}_{d}-{\textstyle\frac{1}{3}}h^{ab}h_{cd}\right)\dot{T}^{cd}\ .
\end{split}
\end{equation}
Note that, in general, the time derivative of vectors and tensors does not commute with the projection of these quantities on the spatial hypersurfaces according to,
\begin{equation}
\begin{split}
&\dot{V}^{\langle a\rangle} \neq({V^{\langle a\rangle}})^{\cdot}\ ,\\
&\dot{T}^{\langle ab\rangle} \neq({T^{\langle ab\rangle}})^{\cdot}\ .
\end{split}
\end{equation}
The projection of the covariant time derivative of a quantity ${Q^{a\dots}}_{b\dots}$ on the spatial hypersurfaces is defined as,
\begin{equation}
^{(3)}\left({Q^{a\dots}}_{b\dots}\right)^{\cdot}\equiv{h^{a}}_{c}\dots {h^{d}}_{b}\dots u^{e}\nabla_{e}{Q^{c\dots}}_{d\dots}\ .
\end{equation}
It is also useful to define the projected covariant curl as,
\begin{equation}
\curl Q_{a\dots b}\equiv\varepsilon_{cd\langle a}D^{c}{Q^{d}}_{\dots b\rangle}\ .\label{curl}
\end{equation}

Information relating to the kinematics is contained in the covariant derivative of $u^{a}$ which can be split into irreducible parts, defined by their symmetry properties,
\begin{align}
\begin{split}
\nabla_{a}u_{b}&=-u_{a}a_{b}+D_{a}u_{b}\\
&=-u_{a}a_{b}+{\textstyle\frac{1}{3}}\Theta h_{ab}+\sigma_{ab}+\omega_{ab}\ ,\label{kinematics}
\end{split}
\end{align}
where
\begin{itemize}
\item $a^{a}\equiv u^{b}\nabla_{b}u^{a}$ is the relativistic acceleration vector, representing the degree to which matter moves under forces other than gravity.
\item $\Theta\equiv D_{a}u^{a}$ is the scalar describing the volume rate of expansion of the fluid (with $H={\textstyle\frac{1}{3}}\Theta$ the Hubble parameter).
\item $\sigma_{ab}\equiv D_{\langle a}u_{b\rangle}$ is the trace-free rate-of-shear tensor describing the rate of distortion of the fluid flow.
\item $\omega_{ab}\equiv D_{[a\vphantom]}u_{\vphantom[b]}$ is the antisymmetric vorticity tensor describing the rotation of the fluid relative to a non-rotating frame.
\end{itemize}
These kinematical quantities have the following properties,
\begin{align}
&a_{a}u^{a}=0\ ,\nonumber\\
&\sigma_{ab}u^{b}=0\ ,\ \ \sigma_{ab}=\sigma_{(ab)}\ ,\ \ {\sigma^{a}}_{a}=0\ ,\\
&\omega_{ab}u^{b}=0\ ,\ \ \omega_{ab}=\omega_{[ab]}\ ,\ \ {\omega^{a}}_{a}=0\ .\nonumber
\end{align}

Note that in presence of vorticity the spatially projected covariant derivatives do not commute, which implies that any scalar field $S$ has to satisfy the non-commutation relation,
\begin{equation}
D_{[a\vphantom]}D_{\vphantom[b]}S=\omega_{ab}\dot{S}\ .\label{kin non com}
\end{equation}

It is useful to introduce a vorticity pseudovector $\omega^{a}$, which is the dual of the vorticity tensor $\omega_{bc}$ and is defined as,
\begin{equation}
\omega^{a}\equiv{\textstyle\frac{1}{2}}\varepsilon^{abc}\omega_{bc}\ ,
\end{equation}
and a vorticity scalar given by,
\begin{equation}
\omega=\left(\omega_{a}\omega^{a}\right)^{1/2}\ .\label{Vorticity scal}
\end{equation}

To build a covariant linear perturbation theory, we will linearise the quantities of the $1+3$ covariant formalism about an FLRW background space-time which is, by definition, homogeneous and isotropic. Gauge invariance of the perturbed quantities is guaranteed by the Gauge Invariance Lemma, which states that if a quantity vanishes in the background space-time, then it is gauge invariant at first-order~\cite{Stewart:1974}. The homogeneity of the background space-time implies that the acceleration $a^{a}$ is a first-order variable since it vanishes in the background space-time (i.e. at zeroth order). Similarly, the isotropy of the background space-time implies that rate of shear $\sigma_{ab}$ is also a first-order variable. Finally, the existence of hypersurfaces orthogonal to the worldline in the background space-time implies that vorticity tensor $\omega_{ab}$ and the vorticity covector $\omega_{a}$ are first-order variables. Therefore, the only zero-order kinematic quantity is the rate of expansion $\Theta$.

In the background space-time, which means to zero-order in the dynamical variables, the covariant derivative of the worldline thus becomes,
\begin{equation}
\nabla_{a}u_{b}={\textstyle\frac{1}{3}}\Theta h_{ab}\ .\label{zero kinematics}
\end{equation}

It is also useful to define (up to some constant factor) a zero-order scale factor $R$ such that,
\begin{equation}
\Theta=3H\equiv\frac{\dot{R}}{R}\ ,\label{scale factor}
\end{equation}
where $H$ is the cosmic Hubble scale factor.

Finally, it is also convenient to introduce a conformal time variable $\hat{\tau}$ satisfying the differential relation,
\begin{equation}
d\hat{\tau}\equiv\frac{d\tau}{R}\ ,\label{conformal time}
\end{equation}
which implies that the derivatives with respect to cosmic time $\tau$ and conformal time $\hat{\tau}$ of a quantity ${Q^{a\cdots}}_{b\cdots}$ are related by,
\begin{equation}
{Q^{\prime\,a\cdots}}_{b\cdots}=R{\dot{Q}^{a\cdots}}_{\phantom{a\cdots}b\cdots}\ ,\label{derivatives relation}
\end{equation}
where a prime denotes a derivative with respect to conformal time $\hat{\tau}$. The conformal Hubble scale factor $\mathcal{H}$ is related to the cosmic scale factor $H$ by $\mathcal{H}=RH$.

\section{First-order adiabatic dynamics of a perfect fluid about an FLRW background}
\label{First-order adiabatic dynamics of a perfect fluid about an FRW background}

We will now use the $1+3$ covariant formalism, outlined in Sec.~\ref{Kinematics in the 1+3 covariant formalism}, to describe the adiabatic dynamics of a perfect fluid to first-order in the dynamical variables, which means neglecting second- and higher-order products of first-order dynamical variables. We then perform a perturbation analysis of such a fluid about an FLRW background, which will be used in Sec.~\ref{Scalar perturbations}-\ref{Tensor perturbations} to describe scalar, vector and tensor perturbations respectively. Thus, we require the cosmological fluid to be highly symmetric on large scales but allow for generic inhomogeneities on small scales.

The dynamics of a perfect fluid is described by the Einstein field equations, which read,
\begin{eqnarray}
R_{ab}-{\textstyle\frac{1}{2}}g_{ab}\mathcal{R}=\kappa T_{ab}\ ,\label{Einstein eq} 
\end{eqnarray}
where $R_{ab}$ and $\mathcal{R}$ are respectively the Ricci tensor and scalar. The dynamical model is fully determined by the matter content and the curvature. The matter content is described by the stress-energy momentum tensor $T_{ab}$. For a perfect fluid, using the $1+3$ formalism, it can be recast as,
\begin{eqnarray}
T_{ab}=\rho u_{a}u_{b}+ph_{ab}\ ,\label{stress energy mom 1+3}
\end{eqnarray}
where $\rho$ is the energy density and $p$ the pressure of the fluid. We assume the fluid to be a specific linear barotropic fluid so that it satisfies the equation-of-state,
\begin{eqnarray}
p=w\rho\ ,\label{eq of state}
\end{eqnarray}
where $w$ is the equation-of-state parameter. The energy density $\rho$ and the pressure $p$ are zero-order variables that do not vanish on the background. 

For an adiabatic flow, the speed of sound is defined as
\begin{equation}
c_s^2\equiv\frac{dp}{d\rho}\ ,\label{speed of sound}
\end{equation}
and the time derivative of the equation-of-state parameter satisfies,
\begin{equation}
\dot{w}=-\Theta(c_s^2-w)(1+w)\ ,\label{deriv w}
\end{equation}
and is derived using the energy conservation equation~\eqref{En cons
eq}. Note that from~\eqref{deriv w}, it follows that $c_s^2=w$ if $\dot{w}=0$. 

All the information related to the curvature is encoded in the Riemann tensor which can be decomposed as~\cite{Hawking:1966},
\begin{equation}
{R^{ab}}_{cd}={C^{ab}}_{cd}+2{\delta^{[a\vphantom]}}_{[c\vphantom]}{R^{\vphantom[b]}}_{\vphantom[d]}- {\textstyle\frac{1}{3}}\mathcal{R}{\delta^{a}}_{[c\vphantom]}{\delta^{b}}_{\vphantom[d]}\ ,\label{Riemann tens}
\end{equation}
where ${C^{ab}}_{cd}$ is the Weyl tensor constructed to be the trace-free part of the Riemann tensor.

By analogy to classical electrodynamics, the Weyl tensor itself can be split, relative to the worldline $u^{a}$, into an `electric' and a `magnetic' part~\cite{Hawking:1966} according to,
\begin{align}
&E_{ab} = C_{acbd}u^{c}u^{d}\ ,\label{Elec}\\
&H_{ab} = \,^{\ast}C_{acbd}u^{c}u^{d} =
{\textstyle\frac{1}{2}}\varepsilon_{ade}{C^{de}}_{bc}u^{c}\ ,\label{Magn}
\end{align}
where $\vphantom{}^{\ast}C_{abcd}$ is the dual of the Weyl tensor. Their properties follow directly
from the symmetries of the Weyl tensor,
\begin{equation}
\begin{split}
&E_{ab}u^{b}=0\ ,\ \ E_{ab}=E_{(ab)}\ ,\ \ \, {E^{a}}_{a}= 0\ ,\\
&H_{ab}u^{b}=0\ ,\ \ H_{ab}=H_{(ab)}\ ,\ \ {H^{a}}_{a}= 0\ .\label{Electric magnetic prop}
\end{split}
\end{equation}
These electric and magnetic parts of the Weyl tensor represent the `free gravitational field', enabling gravitational action at a distance and describing tidal forces and gravitational waves. The Weyl tensor vanishes on a conformally flat background space-time such as FLRW models. Thus, the electric $E_{ab}$ and magnetic part $H_{ab}$ of the Weyl tensor are first-order variables.

The Ricci tensor $R_{ab}$ is simply obtained by substituting the expression~\eqref{stress energy mom 1+3} for the stress energy momentum tensor into the Einstein field equations~\eqref{Einstein eq},
\begin{eqnarray}
R_{ab}={\textstyle\frac{\kappa}{2}}\left(\rho+3p\right)u_{a}u_{b}
+{\textstyle\frac{\kappa}{2}}\left(\rho-p\right)h_{ab}\ .\label{Ef Ricci tensor}
\end{eqnarray}

The Riemann tensor $R_{abcd}$ can now be recast in terms of the Ricci tensor~\eqref{Ef Ricci tensor}, the electric~\eqref{Elec} and magnetic~\eqref{Magn} parts of the Weyl tensor according to the decomposition~\eqref{Riemann tens} in the following way,  
\begin{equation}
\begin{split}
{R^{ab}}_{cd}\ = &{\textstyle\frac{2}{3}}\kappa\left(\rho + 3p\right)u^{[a\vphantom]}u_{[c\vphantom]}{h^{\vphantom[b]}}_{\vphantom[d]}
	+ {\textstyle\frac{2}{3}}\kappa\rho {h^{a}}_{[c\vphantom]}{h^{b}}_{\vphantom[d]}\\
\phantom{{R^{ab}}_{cd}\ = }&+4u^{[a\vphantom]}u_{[c\vphantom]}{E^{\vphantom[b]}}_{\vphantom[d]} + 4{h^{[a\vphantom]}}_{[c\vphantom]}{E^{\vphantom[b]}}_{\vphantom[d]}\\
\phantom{{R^{ab}}_{cd}\ = }&+2\varepsilon^{abe}u_{[c\vphantom]}H_{\vphantom[d]e} + 2\varepsilon_{cde}u^{[a\vphantom]}H^{\vphantom[b]e}\ .\label{Riemann tensor}
\end{split}
\end{equation}
The Riemann tensor restricted to the spatial hypersurface, ${^{(3)}R^{ab}}_{cd}$, is related to the Riemann tensor defined on the whole space-time, ${R^{ab}}_{cd}$, by
\begin{eqnarray}
^{(3)}{R^{ab}}_{cd}={h^{a}}_{e}{h^{b}}_{f}{h^{g}}_{c}{h^{h}}_{d}{R^{ef}}_{gh}-2{v^{a}}_{[c\vphantom]}{v^{b}}_{\vphantom[d]}\ .\label{Riemann tensor 3} 
\end{eqnarray}
where the tensor $v_{ab}$ is defined as,
\begin{equation}
v_{ab}\equiv D_{a}u_{b}\ .\label{Expansion tensor}
\end{equation} 
For a perfect fluid, using the decomposition~\eqref{Riemann tensor}, the spatial Riemann tensor~\eqref{Riemann tensor 3} becomes,
\begin{equation}
^{(3)}{R^{ab}}_{cd}={\textstyle\frac{2}{3}}\kappa\rho{h^{a}}_{[c\vphantom]}{h^{b}}_{\vphantom[d]}+4{h^{a}}_{[c\vphantom]}{E^{b}}_{\vphantom[d]}-2{v^{a}}_{[c\vphantom]}{v^{b}}_{\vphantom[d]}\ .\label{3 Riemann tensor}
\end{equation}
To zero-order, the decomposition of the spatially projected Riemann tensor~\eqref{Riemann tensor 3} reduces to,
\begin{equation}
^{(3)}{R^{ab}}_{cd}=\frac{2K}{R^2}{h^{a}}_{[c\vphantom]}{h^{b}}_{\vphantom[d]}\ .\label{Riemann 3 zero}
\end{equation}
where $K$ is the curvature parameter, which is related to the Gaussian curvature $\mathcal{K}$ by,
\begin{equation}
{^{(3)}R}=\mathcal{K}=\frac{6K}{R^2}\ ,\label{Gaussian curvature}
\end{equation}
where $^{(3)}\mathcal{R}$ is the spatial curvature scalar, which is obtained by twice contracting the spatially projected Riemann tensor~\eqref{Riemann tensor 3}.

In general, there are three sets of dynamical equations for a perfect fluid. These sets are derived, respectively, from the Ricci identities, the Bianchi identities, once- and twice-contracted. We present now each set in turn and expand the dynamical equations to first-order about an FLRW background space-time.

\subsection{Ricci identities}

The first set of dynamical equations arises from the Ricci identities. These identities can firstly be applied to the whole space-time and secondly to the spatial hypersurfaces according to,
\begin{align}
&\nabla_{[a\vphantom]}\nabla_{\vphantom[b]}u_{c}={\textstyle\frac{1}{2}}R_{abcd}u^{d}\ ,\label{Ricci identities}\\
&D_{[a\vphantom]}D_{\vphantom[b]}v_{c}={\textstyle\frac{1}{2}}{\vphantom{a}^{(3)}}R_{abcd}v^{d}\ ,\label{Projected Ricci}
\end{align}
where the spatial vectors $v^{a}$ are orthogonal to the worldline, i.e. $v^{a}u_{a}=0$. 

The information contained in the Ricci identities~\eqref{Ricci identities}-\eqref{Projected Ricci} can be extracted by projecting them on different hypersurfaces using the decomposition of the corresponding Riemann tensors~\eqref{Riemann tensor}-\eqref{3 Riemann tensor} and following the same procedure as in~\cite{Brechet:2007}.

To first-order, the Ricci identities applied to the whole space-time~\eqref{Ricci identities} yield three propagation equations, which are respectively the Raychaudhuri equation, the rate of shear propagation equation and the vorticity propagation equation,
\begin{align}
&\dot{\Theta}=-{\textstyle\frac{1}{3}}\Theta^2-{\textstyle\frac{\kappa}{2}}\left(\rho+3p\right)+D^ba_b\ ,\label{Raychaudhuri eq}\\
&\dot{\omega}_{\langle a\rangle}=-{\textstyle\frac{2}{3}}\Theta\,\omega_{a}+{\textstyle\frac{1}{2}}\curl a_{a}\ ,\label{Vorticity prop eq}\\
&\dot{\sigma}_{\langle ab\rangle}=-{\textstyle\frac{2}{3}}\Theta\,\sigma_{ab}-E_{ab}+D_{\langle a}a_{b\rangle}\ ,\label{Rate shear prop eq}
\end{align}
and three constraint equations,
\begin{align}
&D^{a}\omega_{a}=0\ ,\label{Constr eq 0}\\
&D^{b}\sigma_{ab}={\textstyle\frac{2}{3}}D_{a}\Theta-\curl\omega_{a}\ ,\label{Constr eq 1}\\
&H_{ab}=\curl\sigma_{ab}-D_{\langle a}\omega_{b\rangle}\ .\label{Constr eq 2}
\end{align} 

The Ricci identities applied to the spatial hypersurfaces express the spatial curvature. Their contractions yield the spatial Ricci tensor $^{(3)}R_{ab}$ and scalar $^{(3)}\mathcal{R}$ respectively, which to first-order are given by,
\begin{align}
&^{(3)}R_{ab}={\textstyle\frac{1}{3}}^{(3)}\mathcal{R}{h_{ab}}-{\textstyle\frac{1}{3}}\Theta\left(\sigma_{ab}-\omega_{ab}\right)+E_{ab}\ ,\label{Ricci tensor eq}\\
&\mathcal{K}=-{\textstyle\frac{2}{3}}\Theta^2+2\kappa\rho\ .\label{Gauss Codacci}
\end{align}
To zero-order, using the expression for the Gaussian curvature~\eqref{Gaussian curvature}, the contractions of the spatial Ricci identities~\eqref{Ricci tensor eq} and~\eqref{Gauss Codacci} reduce respectively to,
\begin{align}
&^{(3)}R_{ab}={\textstyle\frac{1}{3}}^{(3)}\mathcal{R}{h_{ab}}\ ,\label{Ricci tens first}\\
&\frac{1}{9}\Theta^2=\frac{\kappa}{3}\rho-\frac{K}{R^2}\ .\label{Friedmann}
\end{align}
Note that the above expression~\eqref{Friedmann} is the Friedmann equation. It is useful to recast the Friedmann~\eqref{Friedmann} and Raychaudhuri~\eqref{Raychaudhuri eq} equations in terms of conformal time. To zero-order, these equations are respectively given by,
\begin{align}
&\mathcal{H}^2={\textstyle\frac{\kappa}{3}}\rho R^2-K\ ,\label{Friedmann 0}\\
&\mathcal{H}^{\prime}=-{\textstyle\frac{\kappa}{6}}\rho R^2(1+3w)\ .\label{Raychaudhuri 0}
\end{align}
Finally, it is convenient to recast them as,
\begin{align}
&\mathcal{H}^2-\mathcal{H}^{\prime}+K={\textstyle\frac{\kappa}{2}}\rho R^2\left(1+w\right)\ ,\label{Conformal 1}\\
&\mathcal{H}^{\prime}=-{\textstyle\frac{\kappa}{2}}\left(1+3w\right)(\mathcal{H}^2+K)\ .\label{Conformal 2}
\end{align}

\subsection{Once-contracted Bianchi identities}
\label{Once-contracted Bianchi identities}

The second and third set of dynamical equations are contained in the Bianchi identities. The Riemann tensor satisfies the Bianchi identities as follows,
\begin{equation}
\nabla^{[e\vphantom]}{R^{\vphantom[ab]}}_{cd}=0 \
.\label{Riemann Bianchi}
\end{equation}
By substituting the expression for the Riemann tensor decomposition~\eqref{Riemann tens} and the effective Einstein field equations~\eqref{Einstein eq} into the Bianchi identities~\eqref{Riemann Bianchi} and contracting two indices ($d=e$), the once-contracted Bianchi identities are found to be,
\begin{equation}
\nabla^{d}{C^{ab}}_{cd}+\nabla^{[a\vphantom]}{R^{\vphantom[b]}}_{c}+{\textstyle\frac{1}{6}}{\delta_{c}}^{[a\vphantom]}\nabla^{\vphantom[b]}\mathcal{R}=0\ .\label{Simple Bianchi}
\end{equation}
In a similar manner to the Ricci identities, the information stored in the once-contracted Bianchi identities has to be projected along the worldlines $u^{a}$ and on the spatial hypersurfaces ${h^{a}}_{b}$.

To first order, the once-contracted Bianchi identities~\eqref{Simple Bianchi} yield two propagation equations, which are respectively the electric and magnetic propagation equations,
\begin{align}
&\dot{E}_{\langle ab\rangle}=-\Theta E_{ab}+\curl H_{ab}-{\textstyle\frac{\kappa}{2}}\left(\rho+p\right)\sigma_{ab}\ ,\label{Electric prop eq}\\
&\dot{H}_{\langle ab\rangle}=-\Theta H_{ab}-\curl E_{ab}\ ,\label{Magnetic prop eq}
\end{align}
and two constraint equations,
\begin{align}
&D^{b}E_{ab}={\textstyle\frac{\kappa}{3}}D_{a}\rho\ ,\label{Constr eq 3}\\
&D^{b}H_{ab}=-\kappa\left(\rho+p\right)\omega_{a}\ .\label{Constr eq 4}
\end{align}

\subsection{Twice-contracted Bianchi identities}
\label{Twice-contracted Bianchi identities}

The third set of equations is given by the twice-contracted Bianchi identities which represent the conservation of the effective stress energy momentum tensor. They are obtained by performing a second contraction ($b=c$) on the once-contracted Bianchi identities~\eqref{Simple Bianchi},
\begin{equation}
\nabla^{b}\left(R_{ab}+{\textstyle\frac{1}{2}}g_{ab}\mathcal{R}\right)=\kappa\nabla^{b}T_{ab}=0\ .\label{Twice Bianchi}
\end{equation}
To first order, the twice-contracted Bianchi identities~\eqref{Twice Bianchi} yield one propagation equations, which is the energy conservation equation,
\begin{equation}
\dot{\rho} = -\Theta\,\left(\rho+p\right)\ ,\label{En cons
eq}
\end{equation}
and one constraint equation, which is the momentum conservation equation,
\begin{equation}
D_{a}p=-a_{a}\left(\rho+p\right)\ .\label{Mom cons eq}
\end{equation}

\section{Scalar perturbations}
\label{Scalar perturbations}

\subsection{Bardeen equation}

Physically, scalar perturbations represent spatial variations of zero-order scalar quantities. Thus, spatial Laplacians of zero-order scalars are natural candidates to describe such perturbations. Since we are interested in determining the time evolution of scalar perturbations in a comoving frame, we choose comoving spatial Laplacians of zero-order scalars as scalar perturbations variables. For the linearised dynamics of a perfect fluid about a homogeneous and isotropic background, there are only four zero-order scalars, the energy density $\rho$, the pressure $p$, the expansion rate $\Theta$ and the Gaussian curvature $\mathcal{K}$. For an adiabatic flow, the pressure is a function of the energy density~\eqref{eq of state}. Thus, to describe the dynamics of adiabatic scalar perturbations, we define three scalar perturbation variables, which are respectively the comoving spatial Laplacian of the energy density $\Phi$, the comoving spatial Laplacian of the expansion rate $\Psi$ and the comoving spatial Laplacian of the Gaussian curvature $\chi$,
\begin{align}
&\Phi\equiv R\kappa\Delta\rho\ ,\label{phi}\\
&\Psi\equiv R\Delta\Theta\ ,\label{psi}\\
&\chi\equiv{\textstyle\frac{1}{2}}R\Delta\mathcal{K}\ ,\label{chi 0}
\end{align}
where the comoving spatial Laplacian $\Delta$ is related to the cosmic spatial Laplacian $D^2$ by,
\begin{equation}
\Delta=R^2D^2=R^2D^{a}D_{a}\ ,\label{spatial com Laplacian}
\end{equation}
and the factor of a half in~\eqref{chi 0} is included to be consistent with the usual definition of the corresponding variable in the `background-based' approach~\cite{Mukhanov:1992}. Similar definitions for the comoving spatial Laplacian of the energy density and the comoving spatial Laplacian of the Gaussian curvature were used respectively by Woszczyna \& Kulak~\cite{Woszczyna:1989} and Bruni, Dunsby \& Ellis~\cite{Bruni:1992}. For a vanishing background curvature, it is worth mentioning that the curvature perturbation does not vanish, since it is a first-order variable. The dynamics of the scalar perturbations is obtained by taking the comoving spatial Laplacian of the scalar propagation equations~\eqref{En cons eq},~\eqref{Raychaudhuri eq},~\eqref{Gauss Codacci} and the spatial gradient of the constraint~\eqref{Mom cons eq} in order to express the constraint in terms of a comoving spacial Laplacian of a zero-order scalar. The dynamical equations of the scalar perturbations are respectively the comoving spatial Laplacian of the energy conservation equation~\eqref{En cons eq}, the comoving spatial Laplacian of the Raychaudhuri equation~\eqref{Raychaudhuri eq}, the comoving spatial Laplacian of the Gauss-Codacci equation~\eqref{Gauss Codacci} and the spatial gradient of the momentum conservation equation, which to first-order reduce to,
\begin{align}
&\Delta\dot{\rho}+\Theta\Delta(\rho+p)+(\rho+p)\Delta\Theta=0\ ,\label{D rho}\\
&\Delta\dot{\Theta}+{\textstyle\frac{2}{3}}\Theta\Delta\Theta+{\textstyle\frac{\kappa}{2}}\Delta(\rho+3p)-\Delta\left(D^{b}a_{b}\right)=0\ ,\label{D Theta}\\
&\Delta\mathcal{K}+{\textstyle\frac{4}{3}}\Theta\Delta\Theta-2\kappa\Delta\rho=0\ ,\label{D K}\\
&\Delta p+R^2\left(\rho+p\right)D^{b}a_{b}=0\ .\label{D p}
\end{align}
In order to reverse the order of the comoving spatial Laplacian and the time derivative of a scalar field $S$ in the propagation equations~\eqref{D rho} and~\eqref{D Theta}, it is useful to introduce the first-order scalar identity,
\begin{equation}
\Delta\dot{S}=\left(\Delta S\right)^{\cdot}-{\textstyle\frac{1}{3}}\Theta\Delta S-R^2\dot{S}D^{b}a_{b}\ .\label{f identity}
\end{equation}  
The dynamical equations~\eqref{D rho},~\eqref{D Theta}, and~\eqref{D K} are recast in terms of the comoving spatial Laplacian~\eqref{phi},~\eqref{psi} and~\eqref{chi 0} using the the scalar identity~\eqref{f identity}, the gradient of the momentum conservation equation~\eqref{D p}, the Friedmann equation~\eqref{Friedmann}, the Raychaudhuri equation~\eqref{Raychaudhuri eq} and the energy conservation equation~\eqref{En cons eq}. To first-order, in a comoving frame, the dynamical equations reduce to,
\begin{align}
&\dot{\Phi}+{\textstyle\frac{1}{3}}\Theta\Phi+\kappa\rho(1+w)\Psi=0\ ,\label{t phi}\\
&\dot{\Psi}+\left(\frac{1}{2}+\frac{3Kc_s^2}{R^2\kappa\rho(1+w)}\right)\Phi+\frac{c_s^2}{R^2\kappa\rho(1+w)}\Delta\Phi=0\ ,\label{t Psi}\\
&\chi+{\textstyle\frac{2}{3}}\Theta\Psi-\Phi=0\ .\label{t chi}
\end{align}
In order to determine the dynamics of the scalar perturbations, it is useful to express the dynamical equations~\eqref{t phi},~\eqref{t Psi} and~\eqref{t chi} in terms of conformal time according to,
\begin{align}
&\Phi^{\prime}+\mathcal{H}\Phi+R\kappa\rho(1+w)\Psi=0\ ,\label{conf phi}\\
&\Psi^{\prime}+\left(\frac{R}{2}+\frac{3Kc_s^2}{R\kappa\rho(1+w)}\right)\Phi+\frac{c_s^2}{R\kappa\rho(1+w)}\Delta\Phi=0\ ,\label{conf Psi}\\
&\chi+\frac{2\mathcal{H}}{R}\Psi-\Phi=0\ .\label{conf chi}
\end{align}

In order to determine the conformal time evolution of the density perturbation variable $\Phi$, the dynamics has to be recast in terms of a second-order differential equation for $\Phi$. By differentiating the $\Phi$-propagation equation~\eqref{conf phi} with respect to conformal time, using the $\Psi$-propagation equation~\eqref{conf Psi} to substitute $\Phi$ for $\Psi$ and the zero-order relations~\eqref{Conformal 1} and~\eqref{Conformal 2}, a second-order differential equation for $\Phi$ is obtained according to,
\begin{equation}
\begin{split}
\Phi^{\prime\prime}&+3\mathcal{H}\left(1+c_s^2\right)\Phi^{\prime}\\
&+\left[(1+3c_s^2)(\mathcal{H}^2-K)+2\mathcal{H}^{\prime}\right]\Phi-c_s^2\Delta\Phi=0\ ,\label{Bardeen eq}
\end{split}
\end{equation}
which is the Bardeen equation, denoted [4.9] in~\cite{Bardeen:1980} and identified by Woszczyna \& Kulak~\cite{Woszczyna:1989} in the $1+3$ covariant formalism. 

We now formally relate the energy density perturbation variable $\Phi$ used in the $1+3$ covariant approach to the Bardeen variables $\Phi_A$ and $\Phi_H$ used in the `background-based' approach. By taking the comoving spatial Laplacian of the divergence of the electric part of the Weyl tensor $E_{ab}$~\eqref{Constr eq 3}, the energy density perturbation $\Phi$ is found to be related to $E_{ab}$ as, 
\begin{equation}
\Phi=3R^3D^aD^bE_{ab}\ .\label{Bardeen Weyl}
\end{equation}
The expression of the electric part of the Weyl tensor $E_{ab}$ in terms of the Bardeen variables $\Phi_A$ and $\Phi_H$ was first established by Bruni, Dunsby \& Ellis~\cite{Bruni:1992} in equations [113-114]. For a perfect fluid (i.e. in absence of anisotropic stress), the Bardeen variables have the same norm but opposite signs (i.e. $\Phi_H=-\Phi_A$). Substituting the relations [113-114] derived by Bruni, Dunsby \& Ellis~\cite{Bruni:1992} into~\eqref{Bardeen Weyl}, the energy density perturbation $\Phi$ is found to be the fourth-order derivative of the Bardeen variable $\Phi_A$ according to,
\begin{equation}
\Phi=\Box\Phi_A\ ,\label{Bardeen Weyl 2}
\end{equation}
where,
\begin{equation}
\Box\equiv 3D^aD^bD_{\langle a}D_{b\rangle}\ .
\end{equation}
Note that by taking spatial derivatives of the electric part of the Weyl tensor, the vector and tensor perturbation terms contained in equations [113-114] vanish. To first-order, the conformal time derivative and the comoving spatial Laplacian commute with the spatial differential operator $\Box$,
\begin{align}
&\Phi^{\prime}=\Box\Phi_A^{\prime}\ ,\\
&\Delta\Phi=\Box\Delta\Phi_A\ .
\end{align}
Thus, to first-order, the dynamics of the energy density perturbation $\Phi$ is identical to the dynamics of the Bardeen variable $\Phi_A$ since the Bardeen equation~\eqref{Bardeen eq} is entirely determined by conformal time derivatives and spatial Laplacians of $\Phi$. Therefore, in the $1+3$ covariant formalism, $\Phi$ is the analog of $\Phi_A$.

\subsection{Mukhanov-Sasaki equation}

The Bardeen equation describes the time evolution of the density perturbation variable $\Phi$. Similarly, the Mukhanov-Sasaki equation describes the time evolution of the curvature perturbation variable, which in the spatially-curved case we will denote by $\zeta$, and which reduces to $\chi$ defined in~\eqref{chi 0} in the spatially-flat case ($K=0$), as we will now show.

To make contact with the standard definition of the curvature perturbation in the `background-based' approach, it is useful to express $\chi$ in terms of $\Phi$ only. By substituting~\eqref{conf phi} into~\eqref{conf chi}, using the Friedmann equation~\eqref{Friedmann 0}, the expression for $\chi$ to first-order becomes,
\begin{equation}
\chi=\frac{2}{3(1+w)}\left(1+\frac{K}{\mathcal{H}^2}\right)^{-1}\left(\Phi+\frac{\Phi^{\prime}}{\mathcal{H}}\right)+\Phi\ ,\label{curvature guess}
\end{equation}
which, for a spatially-flat background space-time ($K=0$), reduces to,
\begin{equation}
\chi=\frac{2}{3(1+w)}\left(\Phi+\frac{\Phi^{\prime}}{\mathcal{H}}\right)+\Phi\ ,\label{curvature perturb}
\end{equation}
which is the standard definition of the curvature perturbation (see Mukhanov~\cite{Mukhanov:2005}, Durrer~\cite{Durrer:2008}). In the spatially-flat case, the conformal time derivative of the curvature perturbation $\chi$ is given by,
\begin{equation}
\chi^{\prime}=\frac{2c_s^2}{3\mathcal{H}(1+w)}\Delta\Phi\ ,\label{curvature pert deriv flat}
\end{equation}
which shows that the curvature perturbation $\chi$ is a conserved quantity on large scales. To show this explicitly a Fourier transform has to be performed (see Durrer~\cite{Durrer:2008}). However, we note that on large scales the comoving spatial Laplacian of the analog of the Bardeen variable is negligible compared to the comoving scale $\mathcal{H}$, thus satisfying $\Delta\Phi\ll\mathcal{H}$. Nonetheless, the curvature perturbation is not conserved on super-Hubble scales for a spatially-curved background space-time~\cite{Bruni:1992}. 

In the spatially-curved case, Bruni, Dunsby \& Ellis~\cite{Bruni:1992} mention that there is a generalised curvature perturbation $\tilde{C}$, which is conserved on large scales and defined up to a constant amplitude. Thus, we define the generalised curvature variable $\zeta$ as
\begin{equation}
\zeta\equiv\frac{R}{2}\left(\Delta\mathcal{K}-\frac{4K}{R^2(1+w)}\frac{\Delta\rho}{\rho}\right)\ ,\label{zeta}
\end{equation}
where $\zeta\equiv{\textstyle\frac{1}{2}}\tilde{C}$ in order for $\zeta$ to reduce to $\chi$ in the spatially-flat case ($K=0$). The generalised curvature perturbation variable $\zeta$ in the spatially-curved case is related to the curvature perturbation $\chi$ in the spatially-flat case by,
\begin{equation}
\zeta=\chi-\frac{2K}{3\mathcal{H}^2\left(1+w\right)}\left(1+\frac{K}{\mathcal{H}^2}\right)^{-1}\Phi\ .\label{zeta chi}
\end{equation}

We now briefly show that $\zeta$ is conserved on large scales. By substituting~\eqref{curvature guess} into~\eqref{zeta chi}, the generalised curvature perturbation $\zeta$ is recast in terms of the analog of the Bardeen variable $\Phi$ only according to,
\begin{equation}
\zeta=\frac{2}{3(1+w)}\left(1+\frac{K}{\mathcal{H}^2}\right)^{-1}\left[\left(1-\frac{K}{\mathcal{H}^2}\right)\Phi+\frac{\Phi^{\prime}}{\mathcal{H}}\right]+\Phi\ .\label{curvature pert gen}
\end{equation}
For a spatially-curved case, by differentiating~\eqref{curvature pert gen} and substituting the Bardeen equation~\eqref{Bardeen eq}, the conformal time derivative of the generalised curvature perturbation $\zeta^{\prime}$ is found to be,
\begin{equation}
\zeta^{\prime}=\frac{2c_s^2}{3\mathcal{H}(1+w)}\left(1+\frac{K}{\mathcal{H}^{2}}\right)^{-1}\Delta\Phi\ ,\label{curvature pert deriv curved}
\end{equation}
which means that the generalised curvature perturbation $\zeta$ is a conserved quantity on super-Hubble scales. To show this explicitly a harmonic decomposition has to be performed. However, similarly to the spatially-flat case, we note that on large scale the comoving spacial Laplacian of the analog of the Bardeen variable is negligible compared to the comoving scale $\mathcal{H}$, thus satisfying $\Delta\Phi\ll\mathcal{H}$.

In order to determine the conformal time evolution of the curvature perturbation variable $\zeta$, the dynamics has to be recast in terms of a second-order differential equation for $\zeta$. By differentiating the $\zeta$-propagation equation~\eqref{curvature pert deriv curved} with respect to time and using the comoving Laplacian of the generalised curvature perturbation $\zeta$~\eqref{curvature pert gen} to substitute $\zeta$ for $\Phi$, a second order differential equation for $\zeta$ is obtained according to,
\begin{equation}
\zeta^{\prime\prime}+2\left(\frac{\mathcal{H}}{2}\left(1-3c_s^2\right)-\frac{\mathcal{H}^{\prime}}{\mathcal{H}}\right)\zeta^{\prime}
-c_s^2\Delta\zeta=0\ .\label{zeta second order}
\end{equation}

It is convenient to introduce a new variable $z$,
\begin{equation}
z\equiv\frac{R^2}{c_s\mathcal{H}}\left(\frac{\kappa}{3}\rho(1+w)\right)^{1/2}\ ,\label{z}
\end{equation}
which allows the $\zeta$-propagation equation~\eqref{zeta second order} to be recast in a more convenient form as,
\begin{equation}
\zeta^{\prime\prime}+2\frac{z^{\prime}}{z}\zeta^{\prime}
-c_s^2\Delta\zeta=0\ .\label{zeta second order 2}
\end{equation}
Note that using the dynamical equation~\eqref{Conformal 1}, $z$ is rewritten as,
\begin{equation}
z=\frac{R^2}{c_s\mathcal{H}}\left(\mathcal{H}^2-\mathcal{H}^{\prime}+K\right)^{1/2}\ ,\label{z bis}
\end{equation}
which corresponds to the variable defined by Mukhanov et al.~\cite{Mukhanov:1992} in equation [$10.43$b].

It is also useful to define another variable,
\begin{equation}
v\equiv z\zeta\ ,\label{v}
\end{equation}
which is the analog of the variable $v$ defined by Mukhanov et al.~\cite{Mukhanov:1992} in equation [$10.61$]. Using $v$, the dynamics of the second-order dynamical equation~\eqref{zeta second order 2} is explicitly recast in terms of a wave equation given by,
\begin{equation}
v^{\prime\prime}-\left(c_s^2\Delta+\frac{z^{\prime\prime}}{z}\right)v=0\ .\label{v second order}
\end{equation}
This wave equation is the Mukhanov-Sasaki equation in a spatially-curved FLRW background space-time denoted [$11.7$] in~\cite{Mukhanov:1992}.

\subsection{Mukhanov-Sasaki variable}

The scalar variable $v=z\zeta$ is the analog of the Mukhanov-Sasaki variable associated to the scalar perturbations~\cite{Mukhanov:1992, Sasaki:1984} (note that in a spatially-flat case, the analog of the Mukhanov-Sasaki variable reduces to $v=z\chi$). In the general case, the homogeneity and isotropy of the spatial hypersurfaces enable us to perform a harmonic decomposition of the scalar perturbation variable $v$, where $v$ is decomposed into components that transform irreducibly under translations and rotations, and evolve independently, as explained in~\cite{Durrer:2008}. The harmonic analysis of the scalar perturbation $v$ consists of a decomposition into eigenfunctions of the comoving spatial Laplacian of $v$ according to,
\begin{equation}
\Delta v_k=-k^2v_k\ ,\label{harmonic scalar}
\end{equation}    
where $k$ is the eigenvalue of the associated harmonic mode and the $k$-index denotes the eigenvector of the mode. The comoving wavenumber $\nu$ of the scalar mode is defined as,
\begin{equation}
\nu^2=k^2+K\ ,\label{comoving wavenumber scal}
\end{equation}
where $K=\{-1,0,1\}$ is normalised. The comoving wavenumber $\nu$ takes continuous values when $K=\{-1,0\}$ and discrete ones for $K=1$. In particular, the regular normalisable eigenmodes have $\nu\geq0$ for flat and hyperbolic spatial hypersurfaces, and an integer satisfying $\nu\geq1$ for spheric hypersurfaces as explained in~\cite{Lyth:1995,Tsagas:2008}. 

To first order, the dynamics of the scalar perturbations can be rewritten as a serie of decoupled harmonic oscillators. Using the harmonic decomposition~\eqref{harmonic scalar} in terms of $k$, the Mukhanov-Sasaki wave equation~\eqref{v second order} in the $k$-mode is given by,
\begin{equation}
{v_k}^{\prime\prime}+\left(c_s^2k^2-\frac{z^{\prime\prime}}{z}\right)v_k=0\ .\label{scal pert eq}
\end{equation}
Note that~\eqref{scal pert eq} corresponds to a simple harmonic oscillator in the $k$-mode, 
\begin{equation}
{v_k}^{\prime\prime}+\omega_k^2v_k=0\ ,\label{scal pert eq II}
\end{equation}
where the conformal time dependent frequency $\omega_k(\hat{\tau})$ is given by,
\begin{equation}
\omega_k=\left(c_s^2k^2-\frac{z^{\prime\prime}}{z}\right)^{1/2}\ .\label{time frequency scal}
\end{equation}

\subsection{Comparison with scalar perturbations of a scalar field in the `background-based' approach}

The dynamics of the early universe is believed to undergo an inflation phase described by a scalar field~\cite{Guth:1981, Linde:1982}. In a `background-based' approach, the scalar field $\phi$ in the `real' space-time is decomposed into a background component $\phi_0$ and a gauge-dependent perturbation $\delta\phi$ according to,
\begin{equation}
\phi(t,\mathbf{x})={\phi}_0(t)+\delta\phi(t,\mathbf{x})\ .\label{scal field}
\end{equation}
According to Weinberg~\cite{Weinberg:2008}, the energy-momentum tensor of an unperturbed scalar field takes the perfect fluid form with an energy density and pressure respectively given by,
\begin{equation}
\begin{split}
&\rho_0={\textstyle\frac{1}{2}}\dot{\phi}_0^2+V(\phi_0)\ ,\label{rho scal}\\
&p_0={\textstyle\frac{1}{2}}\dot{\phi}_0^2-V(\phi_0)\ ,
\end{split}
\end{equation}
where $V(\phi_0)$ is an arbitrary real potential. The initial conditions for inflation needed to perform the quantisation are set as the model emerges from the Big Bang, where the scalar field dynamics is dominated by the kinetic term and satisfies $\dot{\phi}_0\gg V(\phi_0)$. Therefore, at very early times, the scalar field behaves like stiff matter and the equation of state parameter $w$ is related to the speed of sound by $w=c_s^2=1$~\cite{Durrer:2008}. The dynamics of scalar perturbations of a scalar field in a $1+3$ covariant approach was first investigated by Bruni, Ellis \& Dunsby~\cite{Bruni:1992b}. Langlois\& Vernizzi~\cite{Langlois:2007} later generalised the $1+3$ covariant approach to multi-scalar fields. 

To first order, the dynamical equation for the Bardeen variable is given by,
\begin{equation}
\dot{\Phi}_A+H\Phi_A={\textstyle\frac{\kappa}{2}}\dot{\phi}_0\delta\phi\ ,\label{scal field eq 1}
\end{equation}
which corresponds to equation [$10.1.12$] presented by Weinberg~\cite{Weinberg:2008}. By comparing this dynamical relation with the corresponding result~\eqref{conf phi} found in the $1+3$ covariant approach using the zero-order expression for the energy density and pressure of a scalar field~\eqref{rho scal} and by using the correspondence relation~\eqref{Bardeen Weyl 2}, the $1+3$ covariant perturbation variable $\Psi$ is found to be related to the scalar field perturbation $\delta\phi$ by,
\begin{equation}
\Psi=-\Box\left(\frac{\delta\phi}{2\dot{\phi}_0}\right)\ .\label{expansion pert 1}
\end{equation}
The curvature perturbation variables in the flat case $\chi$ and the curved case $\zeta$ can be recast in terms of the Bardeen variable $\Phi_A$ and the scalar field perturbation variable $\delta\phi$ according to,
\begin{equation}
\begin{split}
&\chi=\Box\left(\Phi_A+\mathcal{H}\frac{\delta\phi}{\phi_0^{\prime}}\right)\ ,\\
&\zeta=\Box\left(\left[1-\frac{K}{3\mathcal{H}^2}\left(1+\frac{K}{\mathcal{H}^2}\right)^{-1}\right]\Phi_A+\mathcal{H}\frac{\delta\phi}{\phi_0^{\prime}}\right)\ .\label{curv pert scal}
\end{split}
\end{equation}
The quantisation variable $z$ can also be expressed in terms of the scalar field perturbation according to,
\begin{equation}
z=\left(\frac{\kappa}{3}\right)^{1/2}\frac{R\phi_0^{\prime}}{\mathcal{H}}\ .\label{z scal field}
\end{equation}
Finally, for a scalar field behaving like a stiff fluid (i.e. $c_s=1$), the Mukhanov-Sasaki wave equation in the $k$-mode~\eqref{scal pert eq} reduces to,
\begin{equation}
{v_k}^{\prime\prime}+\left(k^2-\frac{z^{\prime\prime}}{z}\right)v_k=0\ .\label{scal pert eq scal field}
\end{equation}
Thereby, with the identities presented above, we formally related the first-order scalar perturbation variables in the $1+3$ covariant formalism to the corresponding variables in the `background-based' approach.

\section{Vector perturbations}
\label{Vector perturbations}

Vector perturbations are described by spatially projected and divergence-free vectors~\cite{Tsagas:2008}. The only dynamical variable which satisfies these constraints to first-order is the vorticity pseudo-vector $\omega^{a}$. The vanishing divergence of $\omega^{a}$ to first-order can be deduced from~\eqref{Constr eq 0}.

The dynamics of the vorticity covector $\omega_{a}$ is determined by the vorticity propagation equation~\eqref{Vorticity prop eq}. To obtain an explicit evolution equation in terms of the vorticity covector only, the term involving the acceleration has to be recast in terms of the vorticity. Using the momentum conservation equation~\eqref{Mom cons eq} and the kinetic non-commutation identity~\eqref{kin non com}, we find to first-order,
\begin{equation}
\curl a_{a}=2c_s^2\Theta\omega_{a}\ .\label{vorticity acc}
\end{equation}
Thus, to first-order, the vorticity propagation equation~\eqref{Vorticity prop eq} can be recast as,
\begin{equation}
\dot{\omega}_{\langle a\rangle}+{\textstyle\frac{2}{3}}\Theta\left(1-{\textstyle\frac{3}{2}}c_s^2\right)\omega_{a}=0\ ,\label{vorticity cosmic}
\end{equation}
as shown by Hawking~\cite{Hawking:1966}. It is convenient to express this dynamical equation in a comoving frame according to,
\begin{equation}
\omega^{\prime}_{\langle a\rangle}+2\mathcal{H}\left(1-{\textstyle\frac{3}{2}}c_s^2\right)\omega_{a}=0\ .\label{vorticity conformal}
\end{equation}
Using the vorticity contraction identity,
\begin{equation}
{w^2}^{\prime}=2\omega_{\langle a\rangle}^{\prime}\omega^{a}\ ,\label{vorticity contr}
\end{equation}
a first-order propagation equation for the vorticity scalar $w$ is obtained,
\begin{equation}
\omega^{\prime}=-2\mathcal{H}\left(1-{\textstyle\frac{3}{2}}c_s^2\right)\omega\ ,\label{vorticity scal eq}
\end{equation}
and implies that the vorticity scalar scales as~\cite{Hawking:1966, Tsagas:2008},
\begin{equation}
\omega\propto R^{-2+3c_s^2}\ .\label{vorticity scaling}
\end{equation}

The scaling relation~\eqref{vorticity scaling} implies that during the expansion phase ($R^{\prime}>0$), the scalar amplitude of the vorticity $\omega$ decays if $c_s^2<{\textstyle\frac{2}{3}}$. The inflationary scenario is the simplest known generating mechanism for the initial density fluctuations. For an inflaton field in slow-roll ($w=-1$), the vorticity scalar scales as $\omega\propto R^{-5}$. Hence, the value of the vorticity scalar at the end of slow-roll inflation $\omega_f$ is related to the value of the vorticity scalar at the onset of the slow-roll inflation $\omega_i$ by,
\begin{equation}
\frac{\omega_f}{\omega_i}=\mathrm{exp}\left(-5N\right)\ ,\label{vorticity inflation}
\end{equation}
where $N$ is the number of e-fold during the slow-roll phase. Hence, if the vector perturbations were initially significant, they have decayed by a factor $\mathrm{exp}(5N)$ during slow-roll inflation~\eqref{vorticity inflation} and can safely be neglected in a subsequent quantitative perturbation analysis. 

\section{Tensor perturbations}
\label{Tensor perturbations}

Tensor perturbations are described by spatially projected, symmetric, trace-free and transverse second rank tensors, as explained  in~\cite{Tsagas:2008}. To find a suitable tensor to describe such perturbations, it is useful to split the magnetic part of the Weyl tensor $H_{ab}$ into a transverse part denoted $H_{ab}^{(T)}$ and a non-transverse part denoted $H_{ab}^{(V)}$ according to,
\begin{equation}
H_{ab}=H_{ab}^{(T)}+H_{ab}^{(V)}\ .
\end{equation}
where the $^{(T)}$ and $^{(V)}$ indices refer respectively to tensorial and vectorial degrees of freedom. By construction, $H_{ab}^{(T)}$ is divergence-free and satisfies the requirements for a tensor perturbation. For convenience, in this section, we will not use explicitly the $^{(T)}$ index to refer to the transverse part of the magnetic part of the Weyl tensor, since we are only considering tensorial degrees of freedom. Note that in the irrotational case, the magnetic part of the Weyl tensor is divergence-free to first-order $\--$ this can be deduced from the constraint~\eqref{Constr eq 4}. Thus, in that case, the analysis is restricted to the tensorial degrees of freedom only and $H_{ab}=H_{ab}^{(T)}$. It is also worth mentioning that the electric part of the Weyl tensor $E_{ab}$ cannot qualify as a tensor perturbation, since it is not divergence-free in presence of matter $\--$ this can be inferred from the constraint~\eqref{Constr eq 3}.

\subsection{Grishchuk equation}

In order to obtain a second-order differential equation in terms of $H_{ab}$, it is useful to introduce linearised identities found in~\cite{Challinor:2000}. Using the Ricci identities~\eqref{Ricci identities}-\eqref{Projected Ricci}, the expression for the Riemann tensor to zero-order~\eqref{Riemann 3 zero} and the definition of a $\curl$~\eqref{curl}, a symmetric and spatially projected tensor $T_{\langle ab\rangle}$ to first order has to satisfy the geometric linearised identity,
\begin{equation}
^{(3)}(\curl T_{ab})^{\displaystyle\cdot}=\curl\dot{T}_{\langle ab\rangle}-{\textstyle\frac{1}{3}}\Theta\curl T_{ab}\ ,\label{identities GW I}\\
\end{equation}
and if $T_{ab}$ is transverse to first-order,
\begin{equation}
D^bT_{ab}=0\ ,\label{T trans}
\end{equation}
then it also satisfies the identity,
\begin{equation}
\curl(\curl T_{ab})=-D^2T_{ab}+\frac{3K}{R^2}T_{ab}\ .\label{identities GW II}
\end{equation}
By differentiating the magnetic propagation equation~\eqref{Magnetic prop eq}, substituting the electric propagation equation~\eqref{Electric prop eq} and using the linearised identities~\eqref{identities GW I} and~\eqref{identities GW II}, the dynamical equation for $H_{ab}$ to first order reduces to,
\begin{equation}
\ddot{H}_{\langle ab\rangle}+\frac{7}{3}\Theta\dot{H}_{\langle ab\rangle}+2\left((1-w)\kappa\rho-\frac{3K}{R^2}\right)H_{ab}-D^2H_{ab}=0\ ,\label{H eq cosmic}
\end{equation}
which has been established in~\cite{Challinor:2000}. Using the zero-order dynamical relations~\eqref{Friedmann 0} and~\eqref{Raychaudhuri 0}, it is convenient to recast the dynamics of $H_{ab}$~\eqref{H eq cosmic} in terms of conformal time $\hat{\tau}$ according to,
\begin{equation}
H^{\prime\prime}_{\langle ab\rangle}+6\mathcal{H}H^{\prime}_{\langle ab\rangle}-\left(\Delta-2K-8\mathcal{H}^2-4\mathcal{H}^{\prime}\right)H_{ab}=0\ .\label{H eq conformal}
\end{equation}
In order to eliminate the second term on the LHS of~\eqref{H eq conformal} and find a suitable wave equation for the gravitational waves, it is useful to introduce the rescaled magnetic part of the Weyl tensor $\tilde{H}_{ab}$ defined as
\begin{equation}
\tilde{H}_{ab}\equiv R^3H_{ab}\ .\label{H rescaled}
\end{equation}
The propagation equation of the tensor perturbations~\eqref{H eq conformal} can be now elegantly reformulated to first-order in terms of $\tilde{H}_{ab}$ and yields,
\begin{equation}
\tilde{H}_{\langle ab\rangle}^{\prime\prime}-\left(\Delta-2K+\frac{R^{\prime\prime}}{R}\right)\tilde{H}_{ab}=0\ ,\label{GW eq}
\end{equation}
which is our version of the Grishchuk equation~\cite{Grishchuk:1974} describing the dynamics of primordial gravitational waves in a spatially-curved case.

\subsection{Tensor decomposition of $\tilde{H}_{ab}$}

In order to determine the scalar canonical variables associated to the tensor perturbations, which are the scalar amplitudes of the tensor perturbations and will be referred to as canonical variables~\cite{Grishchuk:1974} in the current work, a tensor decomposition of $\tilde{H}_{ab}$ has to be performed. From the properties of the magnetic part of the Weyl tensor $H_{ab}$, we deduce that the key tensor $\tilde{H}_{ab}$ is symmetric, trace-free and transverse to first-order. The transversality of $\tilde{H}_{ab}$ can be expressed as,
\begin{equation}
D^{b}\tilde{H}_{ab}=k^{b}\tilde{H}_{ab}=0\ ,\label{transversality}
\end{equation}
where $k^{b}$ is the spatial wavevector of the transverse gravitational waves satisfying $u^b k_b=0$. For convenience, we now define two vectors $e^{(1)a}$ and $e^{(2)a}$ that provide an orthonormal basis for the two-dimensional spatial hypersurface orthogonal to the propagation direction $k^{a}$ of the gravitational waves, and thus satisfy the following constraints,
\begin{equation}
\begin{split}
&u^{a}e^{(1)}_{a}=u^{a}e^{(2)}_{a}=0\ ,\\
&k^{a}e^{(1)}_{a}=k^{a}e^{(2)}_{a}=0\ ,\\
&e^{(1)a}e^{(1)}_{a}=e^{(2)a}e^{(2)}_{a}=1\ ,\\
&e^{(1)a}e^{(2)}_{a}=0\ .\label{trans vec}
\end{split}
\end{equation}
Note that these two vectors are not uniquely defined, which does not hinder our perturbation analysis since any orthonormal vector basis of the two dimensional hypersurface can be rotated to recover our vector basis $\{e^{(1)a}, e^{(2)a}\}$. 

For an irrotational fluid, the Fermi-Walker transport of the basis vectors $e^{(1)}_{a}$ and $e^{(2)}_{a}$ vanishes~\cite{Weinberg:1972},
\begin{equation}
\begin{split}
&u^b\nabla_b e^{(1)}_a - u_a a^be^{(1)}_{b}=0\ ,\\
&u^b\nabla_b e^{(2)}_a - u_a a^be^{(2)}_{b}=0\ .\label{Fermi Transport}
\end{split}
\end{equation}
Projecting the Fermi-Walker transported basis vectors~\eqref{Fermi Transport} on the spatial hypersurface yields,
\begin{equation}
e^{(1)\prime}_{\langle a\rangle}=e^{(2)\prime}_{\langle a\rangle}=0\ .\label{time deriv basis}
\end{equation}

Taking the comoving spatial Laplacian of the constraints~\eqref{trans vec}, we deduce the following identities,
\begin{equation}
\begin{split}
&e^{(1)a}\Delta e^{(1)}_{a}=e^{(2)a}\Delta e^{(2)}_{a}=0\ ,\\
&e^{(1)a}\Delta e^{(2)}_{a}=-e^{(2)a}\Delta e^{(1)}_{a}\ .\label{Laplacian vec id}
\end{split}
\end{equation}

In order to decompose the tensor perturbation tensor $\tilde{H}_{ab}$ into two polarisation modes, we define two covariant, trace-free and linearly independent polarisation tensors,
\begin{equation}
\begin{split}
&e^{+}_{ab}={\textstyle\frac{1}{2}}\left(e^{(1)}_{a}e^{(1)}_{b}-e^{(2)}_{a}e^{(2)}_{b}\right)\ ,\\
&e^{\times}_{ab}={\textstyle\frac{1}{2}}\left(e^{(1)}_{a}e^{(2)}_{b}+e^{(2)}_{a}e^{(1)}_{b}\right)\ ,\label{poltens}
\end{split}
\end{equation}
which satisfy the orthonormality conditions,
\begin{equation}
\begin{split}
&e^{+}_{ab}e^{+ab}=e^{\times}_{ab}e^{\times ab}=1\ ,\\
&e^{+}_{ab}e^{\times ab}=0\ ,\label{ortho cond}
\end{split}
\end{equation}
and thus form an orthonormal tensor basis for the polarisation modes~\cite{Durrer:2008}. From the identities~\eqref{time deriv basis}, we deduce that the spatially projected time derivatives of the polarisation tensors vanish, 
\begin{eqnarray}
e^{+\prime}_{\langle ab\rangle}=\,e^{\times\prime}_{\langle ab\rangle}=0\ .\label{polar tens const}
\end{eqnarray}
Taking the comoving Laplacian of the orthonormality conditions~\eqref{ortho cond} and using the identities~\eqref{Laplacian vec id}, we deduce the following constraints,
\begin{equation}
\begin{split}
&e^{+}_{ab}\Delta e^{+ab}=e^{\times}_{ab}\Delta e^{\times ab}=0\ ,\\
&e^{+}_{ab}\Delta e^{\times ab}=e^{\times}_{ab}\Delta e^{+ab}=0\ .\label{ortho constr}
\end{split}
\end{equation}

The tensor perturbation variable $\tilde{H}_{ab}$ is spatially projected, transverse, traceless, and therefore can be decomposed into two polarisation modes $\{+, \times\}$ according to,
\begin{equation}
\tilde{H}_{ab}=h^{+}e^{+}_{ab}+h^{\times}e^{\times}_{ab}\ ,\label{decom pol}
\end{equation}
where 
\begin{equation*}
\begin{split}
h^{+}=\tilde{H}^{ab}e^{+}_{ab}\ ,\\
h^{\times}=\tilde{H}^{ab}e^{\times}_{ab}\ ,
\end{split}
\end{equation*}
are the scalar amplitudes of the tensor perturbations. The corresponding decomposition in the `background-based' approach is mentioned by Durrer in~\cite{Durrer:2008}. The linear independence of the polarisation tensors in the first-order perturbation analysis allows us to study separately the dynamics of the two decoupled polarisation modes. To keep the notation compact, we introduce the polarisation mode superscript $\lambda=\{+, \times\}$. 

\subsection{Canonical variables}

Contracting the Grishchuk equation~\eqref{GW eq} with the polarisation basis tensors $e^{\lambda}_{ab}$ and using the identities~\eqref{polar tens const} and~\eqref{ortho constr}, we obtain the Grishchuk equation for the scalar amplitude of the tensor perturbations associated to the polarisation mode $\lambda$,
 \begin{equation}
h^{\lambda\prime\prime}-\left(\Delta-2K+\frac{R^{\prime\prime}}{R}\right)h^{\lambda}=0\ .\label{tens pert eq full}
\end{equation}
where $h^{\lambda}$ are the canonical variables, which describe the scalar amplitude of the tensor perturbations in the polarisation modes $\lambda$. There are two canonical variables $\{ h^{+}, h^{\times}\}$ associated to the polarisation modes of the tensor perturbations. 

In a similar manner than for the scalar perturbations, we perform a harmonic decomposition of the tensors perturbations. The harmonics analysis of the tensor perturbations consists in a decomposition into eigenfunctions of the comoving spatial Laplacian of the scalar amplitude of the tensor perturbation variable $h^{\lambda}$ according to,
\begin{equation}
\Delta h^{\lambda}_k=-k^2 h^{\lambda}_k\ ,\label{harmonic tensor}
\end{equation}    
where $k$ is the eigenvalue of the associated harmonic mode and the suffix $k$ denotes the eigenvector of the mode. 

Finally, to first-order, the dynamics of the tensor perturbations for each polarisation mode $\lambda$ can be rewritten as a serie of decoupled harmonic oscillators. Using the harmonic decomposition of the scalar amplitudes in terms of $k$~\eqref{harmonic tensor}, the evolution equation for the scalar amplitude of the tensor perturbations~\eqref{tens pert eq full} in the $k$-mode satisfies,
\begin{equation}
{h^{\lambda}_{k}}^{\prime\prime}+\left(k^2+2K-\frac{R^{\prime\prime}}{R}\right)h^{\lambda}_{k}=0\ ,\label{tens pert eq}
\end{equation}
for each polarisation $\lambda=\{+, \times\}$. The tensor perturbation amplitudes variables $h^{\lambda}$ are identified as the two scalar canonical variables associated to the tensor perturbations~\cite{Grishchuk:1974}. Note that~\eqref{tens pert eq} corresponds to a simple harmonic oscillator with a polarisation $\lambda$ in the $k$-mode,
\begin{equation}
{h^{\lambda}_{k}}^{\prime\prime}+{\omega^{\lambda}_k}^{\,2}h^{\lambda}_{k}=0\ ,\label{tens pert eq II}
\end{equation}
where the conformal time dependent frequency $\omega^{\lambda}_k(\hat{\tau})$ is given by,
\begin{equation}
\omega^{\lambda}_k=\left(k^2+2K-\frac{R^{\prime\prime}}{R}\right)^{1/2}\ .\label{time frequency tens}
\end{equation}

\section{Conclusion}

We performed a perturbation analysis of an adiabatic perfect fluid to first order using the $1+3$ covariant and gauge-invariant formalism and identified the analog of the Mukhanov-Sasaki variable and the canonical variables needed to quantise the scalar and tensor perturbations respectively about a spatially-curved FLRW background space-time. We also determined the dynamics of the vector perturbations, which does not lead to a second order wave-equation, unlike the scalar and tensor perturbation, but to a first order equation involving the vorticity. In an expanding universe, the vorticity decays provided the sound speed squared satisfies $c_s^2<{\textstyle\frac{2}{3}}$ and the vector perturbations can therefore be neglected in the analysis. 

In cosmological perturbation theory, there are in general six degrees of freedom: two scalars, two vectors and two tensors as mentioned by Bertschinger~\cite{Schaeffer:1996}. However, in the perfect fluid case, the two Bardeen potential are related by $\Phi_A=-\Phi_H$, and thus there is only one remaining scalar degree of freedom. To sum up, to first-order, there is one degree of freedom associated with the scalar analog of the Mukhanov-Sasaki variable $v$, two degrees of freedom associated with the divergence-free vorticity covector $\omega_{a}$ and two degrees of freedom related to the two canonical variables $h^{+}$ and $h^{\times}$ representing the scalar amplitudes of the tensor perturbations. Hence, to first-order, the dynamics of the adiabatic perturbations of a perfect fluid are described by five parameters, as expected.

\begin{acknowledgments}

S.~D.~B. thanks the Isaac Newton Studentship and the Sunburst Fund for their support. The authors also thank Marco Bruni, Stephen Gull, Anthony Challinor and Pierre Dechant for useful discussions.

\end{acknowledgments}

\bibliography{references} 

\end{document}